\documentclass{amsart}
\def\rank{\textup{rank}\,}
\def\t{\vec t}
\def\bibref[#1]{\cite{#1}}
\def\note[#1]{\par\medskip\noindent\textbf{#1:}}
 \newtheorem{theorem}{Theorem}[section]
 \makeatletter
\def\@oddhead{\underline{\hbox to \textwidth{Rank One Conditions for Rectangular Matrices \hfill Gekhtman/Kasman}}}
\let\@evenhead\@oddhead \def\@oddfoot{$\overline{\hbox to
\textwidth{\hfill \bf\thepage\hfill}}$} \let\@evenfoot\@oddfoot
\advance\footskip.5in \makeatother

\def\bibref[#1]{\cite{#1}}

\title{Integrable Systems and Rank One Conditions for Rectangular Matrices}
\author{Michael Gekhtman}
\address{Department of  Mathematics / University of Notre Dame / Notre Dame, IN, USA}
\author{Alex Kasman}
\address{Department of Mathematics / College of Charleston / Charleston, SC, USA}

\begin{document}

\begin{abstract}
We provide a determinantal formula for tau-functions of the KP hierarchy in
terms of \textit{rectangular}, constant matrices $A$, $B$ and $C$ satisfying a rank
one condition.  This result is shown to generalize and unify many previous
results of different authors on constructions of tau-functions for
differential and difference integrable systems from \textit{square} matrices satisfying
rank one conditions.  In particular, it contains as explicit special cases
the formula of Wilson for tau-functions of rational KP solutions in terms
of Calogero-Moser Lax matrices as well as our
previous formula for KP tau functions in terms of almost-intertwining
matrices.
\end{abstract}

\maketitle

\section{Introduction}

In many recent papers by different authors, determinantal formulas have
been used to transform constant \textit{square} matrices satisfying a rank one condition
into tau-functions for integrable systems.  In particular, we recall the
result of G. Wilson \bibref[W2] that $n\times n$ matrices $X$ and $Z$
satisfying the ``almost-canonically conjugate'' condition
\begin{equation}
\rank([X,Z]+I)=1
\label{eqn:almost-canon}
\end{equation}
produce tau-functions for rational solutions to the KP hierarchy by the
formula
\begin{equation}
\tau(\t)=\det\left(X+\sum_{i=1}^{\infty} i t_i
Z^{i-1}\right)\label{eqn:wilsonstau}.
\end{equation}
(See also \bibref[BW1,BW2,ginzburg,Roth] .)

The previous result can be interpreted as a relationship between the KP
hierarchy and the Calogero-Moser particle system and is therefore similar
to the relationship between KdV solitons and the Ruijsenaars-Schneider
particle system \bibref[Braden,RS].  In that case one finds that the Lax
matrices $X$ and $Z$ for this particle system satisfy the rank one
condition
$\rank(XZ+ZX)=1$ and that the formula
\begin{equation}
\tau(\t)=\det\left(\exp(\sum_{i=0}^{\infty} t_{2i+1}Z^{2i+1})
X\exp(\sum_{i=0}^{\infty} t_{2i+1}Z^{2i+1})+I\right)\label{eqn:RStau}
\end{equation}
gives a tau-function for a multi-soliton solution of the KdV hierarchy.

Both soliton and rational solutions were produced by a formula in our
previous paper \bibref[KG] in which it was shown that square matrices $X$, $Y$ and $Z$
satisfying the ``almost-intertwining'' condition
\begin{equation}
\rank(XZ-YX)=1\label{eqn:almost-intertwine}
\end{equation}
produce KP tau-functions through the formula
\begin{equation}
\tau(\t)=\det\left(X\exp(\sum_{i=1}^{\infty}
t_iZ^i)+\exp(\sum_{i=1}^{\infty} t_iY^i)\right)\label{eqn:mytau}.
\end{equation}

Previous results have also demonstrated the usefulness of matrices
satisfying these rank one equations in constructing solutions to difference
equations.  In particular, it was shown in \bibref[rnba] that for matrices
$X$ and $Z$ satisfying \eqref{eqn:almost-canon}, the
eigenvalues $x_i^m$ of the matrix
\begin{equation}
\mathbf{X}(m)=-\eta
X\cdot(\lambda_1-Z)-m\eta(\lambda_2-Z)^{-1}\cdot(\lambda_1-Z)\label{eqn:rnba-X}
\end{equation}
(where $\eta$ and $\lambda_i$ are arbitrarily
selected constants) satisfy the rational nested Bethe Ansatz equations
\begin{equation}
\prod_{k=1}^n
\frac{(x_j^m-x_k^{m-1})(x_j^m-x_k^m+\eta)(x_j^m-x_k^{m+1}-\eta)}{(x_j^m-x_k^{m-\
1}+\eta)(x_j^m-x_k^m-\eta)(x_j^m-x_k^{m+1})}=-1\qquad
\forall 1\leq j\leq n. \label{eqn:rnba}
\end{equation}
Moreover, we announced at NEEDS 2001 (an immediate consequence of \bibref[KG]) that the formula
\begin{equation}
\tau_l^{m,n}=\det\left[X(c_1 - Z)^l(c_2  - Z)^m (c_3 - Z)^n+ (c_1  - Y)^l(c_2  -Y)^m(c_3  -Y)^nW\right]
\label{eqn:HBDE-tau}
\end{equation}
gives a solution to the Hirota Bilinear Difference equation
\begin{equation}
(c_2-c_3)\tau_{l+1}^{m,n}\tau_{l}^{m+1,n+1}-(c_1-c_3)\tau_{l}^{m+1,n}\tau_{l+1}^{m,n+1}+
(c_1-c_2) \tau_{l}^{m,n+1}\tau_{l+1}^{m+1,n}=0\label{eqn:HBDE}
\end{equation}
when $X$, $Y$ and $Z$ satisfy \eqref{eqn:almost-intertwine} (for arbitrary
$n\times n$ matrix $W$).

In addition, two formulas have recently been published for Baker-Akhiezer
functions of the $q$-KP hierarchy in terms of matrices satisfying
$q$-variants of these ``almost'' operator identities.  In particular,
the formula
\begin{equation}
\psi(x,z)=\frac{\det\left(xzI+xqZ-zX-XZ-I\right)}{\det(xI-X)\det(zI+qZ)}e^{xz}
\label{eqn:iliev-psi}
\end{equation}
in the case $\rank(XZ-qZX+I)=1)$ is found in \bibref[Iliev].  Similarly, in
\bibref[NC] one finds the formula
\begin{equation}
\psi(x,z)=\frac{\det(qZX-xZ-zX+zxI)}{\det(ZX-xZ-zX+zxI)}e_q^{xz}
\label{eqn:ChalNij-psi}
\end{equation}
for the case $\rank(XZ-qZX)=1$.
In both cases, the interest in
these particular Baker-Akhiezer functions lies in their bispectrality
(cf. \bibref[BispBook]).

Besides the obvious similarities, some explicit connections have been drawn
between these results.  For example, the formula \eqref{eqn:rnba-X} is
easily derived from \eqref{eqn:wilsonstau} (although this is not the way it
was derived in \bibref[rnba]), formula \eqref{eqn:mytau} is able to
construct the rational solutions like \eqref{eqn:wilsonstau} as well as soliton solutions,
and \eqref{eqn:RStau} can be viewed as a special case of \eqref{eqn:mytau}.
The formulas concerning the $q$-KP hierarchy should be able to be related to the others by the correspondence in \bibref[Iliev] (cf. \bibref[AHvM]) between tau-functions and wave functions of the $q$-KP and KP hierarchies.
However, what is clearly lacking is a general framework in which all of
these different formulas appear as special cases.  It is our goal in this
paper to provide such a framework by generalizing the formulas to the case of \textit{rectangular} matrices.

\section{Main Results}

Let $A,C$ be full rank $n\times N$ matrices and let $B$ be an $N\times N$ matrix
for $N>n$.  
For convenience,
we will  assume the non-degeneracy condition
$\det\left[A  C^{\top}\right]\ne 0$ is satisfied.

Let $L=\textup{row}\,A$ be the subspace
in $\mathbb{C}^N$ spanned by the rows of $A$ and $L^\perp$ be its orthogonal
complement with respect to the standard bilinear form $\langle x,y \rangle =\sum_{k=1}^N x_i y_i$.
Fix a basis $u_1,\ldots, u_{N-n}$ of $L^\perp$
and define $U$ to  be the $(N-n)\times N$ matrix with rows $u_1,\ldots, u_{N-n}$.
As in \bibref[KG], let $g(x)=\sum_{i=1}^{\infty} t_ix^i$ be a
power series in $x$ with coefficients that depend on the time variables
$\t=(t_1,t_2,\ldots)$ of the KP hierarchy.

\begin{theorem}
If
\begin{equation}
\textup{rank}(A B U^{\top})\leq 1,\label{eqn:rankonenew}
\end{equation}
then
\begin{equation}
\tau_l^{m,n}=\det\left[A(c_1I-B)^l(c_2I-B)^m(c_3I-B)^nC^{\top}\right]\label{eqn:taunewHBDE}
\end{equation}
is a solution to the Hirota Bilinear Difference Equation \eqref{eqn:HBDE} and
\begin{equation}
\tau(\t)=\det\left[A e^{g(B)}C^{\top}\right]\label{eqn:taunew}
\end{equation}
 is a tau-function of the KP hierarchy.
\end{theorem}
\begin{proof}
It is known (see, e.g. \bibref[KWZ,Zabrodin]) that to
prove that $\tau(\t)$ is a KP tau-function it is sufficient to prove that
$\tau_l^{m,n}:=\tau(\t-l[c_1^{-1}]-m[c_2^{-1}]-n[c_3^{-1}])$ solves the HBDE for all values of the parameters.\footnote{As usual, the Miwa shift $\t+c[z]$ is defined by
$$
\t+c[z]=(t_1+{c}{z},t_2+\frac{cz^2}{2},t_3+\frac{cz^3}{3},\ldots).
$$}
 Hence, our method will be to prove the second claim above by proving the first.  Moreover, it is important to note that it is suficient to prove that HBDE is satisfied when all of the discrete "times" $l,m,n=0$.  This is because $\tau_l^{m,n}=\hat\tau_{0}^{0,0}$ if $\hat\tau$ is the tau-function corresponding to the same choice of $A$ and $B$ but with a different $C$ (multiplied on the right by the transpose of $(c_1I-B)^l\cdots(c_3-B)^n$).  Since the condition in the claim depends only on a property of $A$ and $B$, it is therefore sufficient to consider the restricted version of the equation
\begin{eqnarray*}
(c_2-c_3)\tau(\t-[c_1^{-1}])\tau(\t-[c_2]^{-1}-[c_3^{-1}])-(c_1-c_3)\tau(\t-[c_2^{-1}])\tau(\t-[c_1^{-1}]-[c_3^{-1}])
\\+(c_1-c_2)\tau(\t-[c_3^{-1}])\tau(\t-[c_1^{-1}]-[c_2^{-1}])&=&0.
\end{eqnarray*}
We will reduce this equation to the following identity, proved in \bibref[KG]:
\begin{equation}
h_1(c_1)h_2(c_2,c_3) - h_1(c_2)h_2(c_1,c_3) + h_1(c_3)h_2(c_1,c_2)\equiv 0,
\label{eqn:h3}
\end{equation}
where
$h_1(c_1) = \det [ c_1 - P ]$, $h_2(c_1,c_2)=\det[(c_1-P)(c_2-P) + Q]$ with
$P,Q$ $n\times n$ matrices and $\textup{rank}(Q)\le 1$.

First, let $V^{\top}$ be any right inverse of $A$ and define
$G=[V^{\top} \ U^{\top}], \hat B = G^{-1} B G$, $\hat
C^{\top}=G^{-1} C^{\top} $ and $M=M(\t)=[I_n \ 0] e^{g(\hat
B)}{\hat C}^{\top}$. Then $ A G = [I_n \ 0]$ and
$\tau(\t)=\det\left[M(\t)\right]$.
Note also that $G^{-1} = \left
[\begin{array}{c} A\\ * \end{array} \right ]$.

The Miwa  shift  $\tau(\t-[c^{-1}])$
can be computed as follows:
$$\tau(\t-[c^{-1}])= \det\left[ [I_n  0] e^{g(\hat
B)} e^{\ln(I_N - c^{-1} \hat B)}{\hat C}^{\top} \right]=
c^{-n} \tau(\t) \det\left [ c - [I_n  0] \hat B e^{g(\hat
B)}{\hat C}^{\top} M(\t)^{-1} \right ]
$$
and, similarly,
$$
\tau(\t-[c_1^{-1}]-[c_2^{-1}])=c_1^{-n} c_2^{-n}\tau(\t)
\det\left [ [I_n \ 0](c_1 -\hat B)(c_2 -\hat B) e^{g(\hat
B)}{\hat C}^{\top} M(\t)^{-1} \right]
$$
Then it
is not hard to check that a left hand side of the bilinear
difference Hirota equation is proportional to the left hand side of
\eqref{eqn:h3}, if one defines $P= [I_n \ 0]\hat B e^{g(\hat
B)}{\hat C}^{\top} M^{-1}$ and $Q=[I_n \ 0](\hat B)^2 e^{g(\hat
B)}{\hat C}^{\top} M^{-1} - P^2$. Hence, it suffices to show that
$\textup{rank}(Q)\le 1$. Rewrite $Q$ as
$$Q = [I_n \ 0]\hat B \left ( I - e^{g(\hat B)}{\hat C}^{\top} M^{-1} [I_n \ 0]\right )
\hat B e^{g(\hat B)}{\hat C}^{\top} M^{-1}$$
and notice that $e^{g(\hat B)}{\hat C}^{\top}=\left[\begin{array}{c}M\\ * \end{array} \right ]$. Therefore,
$$ I - e^{g(\hat B)}{\hat C}^{\top} M^{-1} [I_n \ 0]=\left [ \begin{array}{cc} 0 & 0\\ * & I_{N-n}\end{array} \right ]\ ,$$
which means that
$$[I_n \ 0]\hat B \left ( I - e^{g(\hat B)}{\hat C}^{\top} M^{-1} [I_n \ 0]\right )=\left ( [I_n \ 0]\hat B
\left [ \begin{array}{c} 0\\ I_{N-n}\end{array} \right ]\right ) [* \ I_{N-n}]\ .$$
But the factor in parentheses is equal to $A B U^{\top}$ and, by our assumption, has a rank less or equal
to $1$. Therefore, the same is true for $Q$, which finishes the proof.
\end{proof}

\note[Remark] It is worth emphasizing that the rank-one condition \eqref{eqn:rankonenew}
does not depend on a choice of bases in subspaces $L$ and $L^\perp$ that correspond
to the matrix $A$.

Since we have constructed a solution to the KP hierarchy, one may be
interested in knowing where in the Sato-Segal-Wilson grassmannian \bibref[Sato,SW]
the corresponding solutions lie.  This can be resolved by studying the
associated Baker-Akhiezer function, $\psi(x,z)$.  In particular, one knows
that the point $W\in Gr$ is the subspace spanned by $\psi$ and its $x$-derivatives evaluated
at $x=0$
$$
W=\langle\psi(0,z), \psi_x(0,z),\psi_{xx}(0,z),\ldots\rangle.
$$
We will show below that there exist polynomials $p(z)$ and $q(z)$ such that
$p(z)H_+\subset W\subset q^{-1}(z)H_+$, which shows by definition that $W$
is in the subgrassmannian $Gr^{rat}$ \bibref[cmbis,W].
\begin{theorem}
The stationary Baker-Akhiezer function corresponding to the tau-function given in
\eqref{eqn:taunew} is
\begin{equation}
\psi(x,z)=\frac{\det(Ae^{xB}(zI-B)C)}{z^N\det(Ae^{xB}C)}e^{xz}.\label{eqn:newpsi}
\end{equation}
As a consequence, we are able to determine that this solution corresponds
to a point in the subgrassmannian $Gr^{rat}$ of rank one KP solutions with
rational spectral curves.
\end{theorem}

\begin{proof}
The formula for $\psi(\t,z)$ follows from the well known formula for the
time-dependent wave function
\bibref[SW]:
$$
\psi(\t,z)=\frac{\tau(\t-[z^{-1}])}{\tau(\t)}e^{g(z)}
$$
which simplifies when evaluated at $\t=(x,0,0,0,\ldots)$ to the formula above since the coefficients of
$-[x]$ are precisely the coefficients in the power series expansion of
$\log(1-x)$.

Another way to identify the subspace $W$ is by its duality with the dual
Baker-Akhiezer function $\psi^*$ which by similar arguments to above can be shown to
have the property that $p(z)\psi^*$ is nonsingular in $z$ for
$p(z)=\det(zI-B)$.  Hence, the innerproduct of any polynomial in $p(z)H_+$
with $\psi^*$ (computed as the path integral of the product around $S^1$) is zero.
Consequently, $p(z)H_+\subset W$ and we see that $W$ is in $Gr^{rat}$.
\end{proof}

It has frequently been found to be useful in the case of solutions corresponding to rational spectral curves \bibref[cmbis,nkdv,SW,W] to identify a solution instead by the finite dimensional space of finitely supported distributions in $z$ that annihilate the Baker-Akhiezer function.
A consequence of the role of the characteristic polynomial $p(z)$ in the
proof above is that the finitely supported distributions in $z$ which
annihilate $z^N\psi(x,z)$ are supported at the eigenvalues of $B$ with
highest derivative taken bounded by the algebraic multiplicity of the
eigenvalue.

\section{Special Cases}

As a corollary to the main theorem above, we can determine the following generalization of our theorem from \bibref[KG]:
\begin{theorem}
Let $X$ be an $n\times(N-n)$ matrix, $Y$ be an $n\times n$ matrix and $Z$ be an $(N-n)\times(N-n)$ matrix such that
$$
\rank(XZ-YX)\leq 1
$$
then \eqref{eqn:taunew} is a tau-function of the KP hierarchy where $A=[X \ I_n]$, $B$ is the block diagonal matrix $B=\textup{diag}[Z,Y]$ and $C$ is an arbitrary full rank $n\times (N-n)$ matrix.
\end{theorem}

To see that this is so, note that
$U$ in \eqref{eqn:rankonenew} can be chosen to be $[-I_n \ X^{\top}]$.
One then finds that condition \eqref{eqn:rankonenew} reads $\rank (X Z - Y X)\le 1$.
This generalizes \eqref{eqn:almost-intertwine}, since there it was assumed that $N=2n$ and $X$ is a square
matrix. In the latter case, one can define $C$ to be $C=[I_n\ I_n]$, which transforms
\eqref{eqn:taunew} into a tau-function $\tau(\t)= \det\left [ X e^{g(Z)} +  e^{g(Y)}\right ]$ that
coincides with \eqref{eqn:mytau}.

In fact, combining the discussion of the previous paragraph with the observation that the distributions annihilating the Baker-Akhiezer function are supported at the eigenvalues of $B$ with order bounded by the algebraic multiplicity confirms the conjecture in \bibref[KG] that the same was true for the eigenvalues of $Y$ and $Z$. 

Moreover, one can also rederive Wilson's formula \eqref{eqn:wilsonstau} and the "almost-canonically conjugate" rank one condition \eqref{eqn:almost-canon} as a special case of our main result.
Consider the case when $N=2n$, $A, C$ and $U$ are defined as in above but $B$ is
chosen in the form
$B=\left [ \begin{array}{cc} Z& 0\\ I_{n} & Z\end{array} \right ]$. Then
$A B U^{\top}= - \left ( [X,Z] + I_n \right )$ and thus,
\eqref{eqn:rankonenew} coincides with \eqref{eqn:almost-canon}. Moreover, in this case
$e^{g(B)}=\left [ \begin{array}{cc} e^{g(Z)}& 0\\ g'(Z) e^{g(Z)} & e^{g(Z)}\end{array} \right ]$,
and the tau-function \eqref{eqn:taunew} becomes
$\tau(\t)= \det [e^{g(Z)}] \det [ X + g'(Z)]$, which is gauge equivalent to the one in
\eqref{eqn:wilsonstau}.

In conclusion,  although many of the details are yet to be fully explored, the formula we have proven above for the first time allows us to consider many different results relating rank one conditions and tau-functions in a unified context.  We plan to address questions of the relationship between the geometry of the space of matrices we utilize and the geometry of the grassmannian, reductions to finite dimensional Hamiltonian (particle) systems, and the case in which $A$, $B$ and $C$ are taken to be infinite dimensional operators in a future paper.

\note[Acknowledgements]  The second author is grateful to the organizers of the conference NEEDS 2001 at which results preliminary to those announced here were presented.  We are also grateful to many people for fruitful and interesting discussions including:
Malcolm Adams,
Yuri Berest,
Harry Braden,
Annalisa Calini,
Pavel Etingof,
Andrew Hone,
Tom Ivey,
Mitch Rothstein,
and Jacek Szmigielski.

\end{document}